# Gigantic-oxidative atomic-layer-by-layer epitaxy for artificially designed complex oxides

Guangdi Zhou[1#], Haoliang Huang[1,2#], Fengzhe Wang[1], Heng Wang[1], Qishuo Yang[1], Zihao Nie[1], Wei Lv[1], Cui Ding[2], Yueying Li[1], Jiayi Lin[1,3], Changming Yue[1], Danfeng Li[2,4], Yujie Sun[1,2], Junhao Lin[1,2], Guang-Ming Zhang[5,6], Qi-Kun Xue[1,2]*, Zhuoyu Chen[1,2]*

[1]Department of Physics and Guangdong Basic Research Center of Excellence for Quantum Science, Southern University of Science and Technology, Shenzhen 518055, China
[2]Quantum Science Center of Guangdong-Hong Kong-Macao Greater Bay Area, Shenzhen 518045, China
[3]Department of Physics, South China University of Technology, Guangzhou 510006, China
[4]Department of Physics, City University of Hong Kong, Kowloon, Hong Kong
[5]State Key Laboratory of Low-Dimensional Quantum Physics, Department of Physics, Tsinghua University, Beijing 100084, China
[6]Frontier Science Center for Quantum Information, Beijing 100084, China

[#]These authors contributed equally.
*E-mail: xueqk@sustech.edu.cn, chenzhuoyu@sustech.edu.cn

**In designing material functionalities for transition metal oxides, lattice structure and *d*-orbital occupancy are key determinants. However, the modulation of these two factors is inherently limited by the need to balance thermodynamic stability, growth kinetics, and stoichiometry precision, particularly for metastable phases. We introduce a methodology, namely the gigantic-oxidative atomic-layer-by-layer epitaxy (GOALL-Epitaxy), enhancing oxidation power 3-4 orders of magnitude beyond conventional pulsed laser deposition (PLD) and oxide molecular beam epitaxy (OMBE), while ensuring atomic-layer-by-layer growth of designed complex structures. Thermodynamic stability is markedly augmented with stronger oxidation at elevated temperatures, whereas growth kinetics is sustained by laser ablation at lower temperatures. We demonstrate the accurate growth of complex nickelates and cuprates, especially an artificially designed structure with alternating single and double $NiO_2$ layers possessing distinct nominal *d*-orbital occupancy, as a parent of high-temperature superconductor. The GOALL-Epitaxy enables material discovery within the vastly broadened growth parameter space.**

In transition metal oxides, complex interplays among charge, spin, orbital, and lattice degrees of freedom, give rise to a rich spectrum of phenomena, such as metal-insulator transitions, magnetism, ferroelectricity, and superconductivity [1,2]. These intertwined orders are rooted in the delicate balance of similar, yet competing and correlated energy scales [3]. For instance, the transition from metal to antiferromagnetic insulator depends on the comparison between the *d*-orbital bandwidth and the *d*-*d* Coulomb interaction. Further complexity arises if the oxygen ligand-to-metal charge transfer energy falls below the Coulomb interaction, shifting the determination to a finer energy scale comparison [4]. This nuanced energy landscape underscores the diverse manifestations of physical properties in transition metal oxides, setting the stage for the design of correlated electron systems.

In designing complex oxides for desired functionalities, two important factors of consideration are the lattice structure and the *d*-orbital occupancy of transition metal ions. The lattice structure not only determines the dimensionality of the active functional layer (e.g. 3D versus 2D), but also governs interfacial coupling between layers within the designed structures [5-9]. Simultaneously, the number of correlated electrons residing in the *d* orbitals at each lattice site directly links to the electronic structure and the Fermi level [10-12]. However, the interplay between the lattice structure and the *d*-orbital occupancy is complex and often interdependent [13,14]. Perturbations of the lattice structures, such as oxygen octahedra rotation [7,15,16] or Jahn-Teller distortion [17], are closely related to the electron count in *d* orbitals due to crystal field anisotropy, illustrating the intricate relationship between these fundamental parameters.

The successful growth of designed complex oxide systems hinges on two critical abilities: (1) the independent control over the intertwined factors of lattice structure and transition metal *d*-orbital occupancy, and (2) the stabilization and manipulation of metastable phases. Achieving these objectives necessitates careful attention to thermodynamic stability, growth kinetics, stoichiometry precision, and the ability to accurately control oxygen content *in situ* over wide ranges. Molecular beam epitaxy (OMBE) and pulsed laser deposition (PLD) stand out as premier techniques for crafting complex oxide epitaxial single-crystalline thin films and heterostructures [18,19]. While these thin film techniques facilitate the creation and exploration of artificial structures and complex physical phenomena, their effectiveness is somewhat diminished for phases that demand substantial oxidation, compared to methods like high-pressure synthesis [20,21]. OMBE, especially when employing alternately shuttered element sources, offers meticulous control over cation stoichiometry and supports atomic-layer-by-layer growth [22-24], whereas PLD is prized for its simplicity, versatility of materials, and capability for higher pressure environments to enhance thermodynamic stability [25,26]. Nonetheless, each technique faces its challenges: OMBE is limited by the vapor pressure of elements and requires a low-pressure environment for the transport of evaporated materials to the substrate, thus constraining its oxidation potential; PLD can lead to stoichiometric imbalances and is less effective for materials with complex and large unit cells, such as Ruddlesden-Popper (RP) phases. Addressing these comprehensive requirements, we present the gigantic-oxidative atomic-layer-by-layer epitaxy (GOALL-Epitaxy) method, detailed subsequently.

Figure 1 illustrates the operational principles of GOALL-Epitaxy with a prominent example growth of an artificially designed nickelate structure as a parent for high-temperature superconductivity [27]. Consist of La, Ni, and O, this structure features the alternating stacking of single- and double-layer $NiO_2$ planes (thus denoted "1212"), with nominal $3d^8$ and $3d^{7.5}$ occupancy, respectively (Figure 1a). The 1212 structure can be regarded as a combination of $La_2NiO_4$ and $La_3Ni_2O_7$ RP phases, the former of which is an antiferromagnetic insulator and the latter of which was recently found to host superconductivity at liquid nitrogen temperatures under high pressure [21]. The 1212 structure does not belong to the RP series and is intrinsically challenging to stabilize thermodynamically, since Ni in $La_2NiO_4$ and $La_3Ni_2O_7$ have different valences. Nonetheless, thanks to the flexibility of GOALL-Epitaxy, the 1212 structure can be well realized in thin film form as shown by the scanning transmission electron microscopy (STEM) image with atom positions identified in Figure 1b.

In general, upon receiving a task of growth, the designed structure is broken down into its constituent oxide atomic layers (LaO and $NiO_2$, in this case, schematized in Figure 1c). Each atomic layer is associated with a specific oxide target (i.e. $LaO_x$ and $NiO_x$). After setting the optimal growth temperature and oxidation conditions, atomic layers are successively deposited onto a chosen substrate through pulsed laser ablation of these oxide targets [18,28]. The number of pulses required for each complete layer, varying from a few tens to several hundreds, is determined by the properties of the target material and the selected laser energy. Layers are deposited in a programmed sequence that reflects the design, with the entire growth process monitored in real-time by reflective high-energy electron diffraction (RHEED), allowing for the immediate characterization of elemental and layer completion information [29], facilitating the atomic-layer-by-layer growth mode [19,28,30-33]. In the example case of 1212 structure, one RHEED oscillation cycle corresponds to one LaO-NiO$_2$-LaO-NiO$_2$-LaO (a double-NiO$_2$ layer) block plus one LaO-NiO$_2$-LaO (a single-NiO$_2$ layer) block, realized by sequentially depositing using LaO$_x$, NiO$_x$, LaO$_x$, NiO$_x$, LaO$_x$, and LaO$_x$, NiO$_x$, LaO$_x$ targets.

Expanding upon atomic-layer-by-layer growth, the GOALL-Epitaxy technique employs purified ozone with a specially designed ozone nozzle (1/2 inch inner diameter) aimed directly at the sample, positioned closely (~4 cm) to establish a highly concentrated oxidation zone close to the substrate (see Figure S1 for simulation analysis). This geometry amplifies oxidation power by around half order of magnitude, a crucial enhancement given the rapid decomposition of ozone molecules upon contact with the heated sample stage. The ozone gas is purified in a liquefaction unit positioned close to the chamber's ozone inlet to minimize decomposition during transport. The RHEED is equipped with a two-stage differentially pump system, enabling atomic-layer-by-layer growth mode to sustain ozone chamber pressures up to 0.1 mbar. In contrast to OMBE, GOALL-Epitaxy synergizes the high-energy plasma plume generated by laser pulses—capable of withstanding high chamber ozone pressures for effective material transport to the substrate. Under strong oxidative conditions, GOALL-Epitaxy's theoretical stoichiometry is as precise as 0.1%, and experimentally demonstrated to be no worse than 0.3% (see Figure S2). Additionally, the chamber pressure can be tuned in accordance with the laser fluence to ensure that ablated particles reach the substrate surface with a suitable kinetic energy, causing neither excessive damage to the grown structure nor insufficient surface diffusion. In the example case of 1212 structure, the oxidative environment was meticulously set to $4 \times 10^{-4}$ mbar ozone plus $9.6 \times 10^{-3}$ mbar oxygen with laser fluence 1.4 J/cm$^2$ for both LaO$_x$ and NiO$_x$ targets. For nominally distinct valences of nickel in different phases, the growth in principle necessitates different oxidative strengths. The mixed oxygen-ozone condition is chosen to accommodate both the growth of single-layer and double-layer structures.

STEM high-angle annular dark field (HAADF) image with a larger field of view (Figure 1d) shows coherent 1212 lattice structure with a sharp substrate/film interface. Atomically resolved energy-dispersive X-ray spectroscopy (EDS) images exhibit element-selective signals (Figure 1e). Within the field of view shown and all other EDS images measured (not shown here), we did not observe obvious interdiffusion of Al and Ni atoms at the interface. On top of the atomically flat substrate, single and double layer of Ni is alternatively stacked, signifying the high-quality realization of the artificially designed complex oxide lattice. X-ray diffraction (XRD, Figure 1f) displays nearly all predicted peaks within measurable range, except for those overlapping with substrate peaks and (00$\underline{24}$) due to lower symmetry. Low-temperature resistivity measurement of our grown film exhibits pure insulating behavior, in contrast to that of bulk crystal [27] featuring a down bending below 140 K.

Figures 2a to 2e illustrate RP phases with varying numbers of transition metal oxide layers—using Ni as an example due to its high oxidation requirements. These variations correspond to different $3d$ orbital occupancy (Figure 2a). Since carrier transport occurs within $NiO_2$ layers, the number of stacking layers controls the effective dimensionality of the electronic system: Between adjacent blocks, $NiO_2$ layers are separated by two insulating LaO layers and the alignment of the oxygen octahedra in the lattice is shifted by approximately one bond length, reducing the probability of carrier hopping. The RHEED oscillations (Figure 2b) reveal that different sequences of ablating La and Ni (i.e., $LaO_x$ and $NiO_x$ targets) yield various structural stackings, with steady oscillation intensity indicating stoichiometric growth (example RHEED patterns shown in Figure S3). The XRD spectrum for each structure (Figure 2c) shows consecutive Bragg peaks along the out-of-plane axis, confirming periodic lattice structures. An STEM image of the two-layer structure $La_3Ni_2O_7$ (Figure 2d) confirms the coherent growth on the $LaAlO_3$ substrate and alternately positioned $LaO$-$NiO_2$-$LaO$-$NiO_2$-$LaO$ blocks. Systematic resistivity-temperature curves for different layer stacking configurations (Figure 2e) reveal that while infinite-layer $LaNiO_3$ displays metallic behavior, structures with 5 to 2 layers exhibit a metal-insulator transition at low temperatures, likely due to reduced dimensionality. The resistivity tends to increase as the number of layers decreases, largely a result of changes of electron occupancy in the $3d$ orbitals, with the single-layer case ($La_2NiO_4$) being highly insulating, aligning with prior findings [32,33].

The enhancements in growth thermodynamics and kinetics provided by GOALL-Epitaxy enable the extension of the growth temperature range for LaNiO$_3$. Specifically, the lower temperature limit is extended to 350 °C under a chamber pressure of $2 \times 10^{-5}$ mbar of O$_3$, while the upper temperature limit reaches 900 °C at a chamber pressure of 0.1 mbar of O$_3$ (see Figure S4). The quality of LaNiO$_3$ growth enables a wide tunable range of in-plane coherent strain reaching up to 4.5% (see Figure S5).

Upon growing the desired complex structures, for independent control of transition metal *d*-orbital occupancy over wide ranges while keeping structural coherence, we implement *in situ* reduction via atomic hydrogen in a dedicated reduction chamber [34-36], as illustrated in Figure S6**a**. The thermally activated atomic hydrogen source is positioned vertically ~20 cm below the sample, to ensure a consistent flux (~$3 \times 10^{15}$ atoms/cm$^2$s) across the sample surface, important for achieving a spatially uniform reduction rate. XRD data presented in Figure S6**b** demonstrate the capability to precisely adjust oxygen content from (La,Sr)NiO$_3$ with approximately $3d^{6.8}$ configuration, to (La,Sr)NiO$_2$, corresponding to around $3d^{8.8}$, by varying atomic hydrogen flux and annealing temperature. Besides atomic hydrogen, reduction through deposition of a thin reductant metal layer, such as evaporated Al from an effusion cell [37], is also possible in our reduction chamber.

The next example showcases the growth of infinite-layer cuprate structures (Figure 3a), with maximized oxidation power of GOALL-Epitaxy. The infinite-layer structure represents the most fundamental parent of cuprate superconductors, characterized by the uninterrupted stacking of CuO$_2$ planes, interspersed with alkaline earth ions [38,39]. Mastering the growth and manipulation of this structure is crucial for designing and creating new cuprate superconductors, although producing high-quality crystals faces significant challenges due to their thermodynamic metastable nature [26,28]. Achieving their growth necessitates exceptionally strong oxidation conditions: a more powerful oxidation environment promotes thermodynamic stability at elevated growth temperatures, thus enhancing growth kinetics and resulting in higher crystalline quality. In previous experiments using OMBE and PLD, the growth temperature typically ranged between 550-600°C due to limited oxidation capabilities [28,40-42]. By utilizing stronger oxidation achievable with GOALL-Epitaxy, stable growth of infinite-layer cuprates can be achieved at temperatures between 600-700 °C, markedly exceeding those of earlier methods, indicating an enhanced thermodynamic stability. Figure 3b shows three example growth process of $CaCuO_2$ (with $SrCuO_2$ buffer), $Sr_{0.5}Ca_{0.5}CuO_2$ (with $SrCuO_2$ buffer), and $SrCuO_2$. Intriguingly, to achieve a 1:1 stoichiometry between Sr and Ca in $Sr_{0.5}Ca_{0.5}CuO_2$, we alternate the deposition of one atomic layer of Sr with one atomic layer of Ca, thus one cycle of RHEED oscillation corresponds to the formation of two unit cells. In Figure 3c, an STEM image demonstrates the coherent $CaCuO_2$ thin film with 20 uc $SrTiO_3$ capping layer and 3 uc $SrCuO_2$ buffer layer grown on the $NdGaO_3$ substrate. The atoms at the interface are clearly visible and consistent with the designed expectations. XRD data (Figure 3d) reveal systematic variations in the out-of-plane lattice constant with different alkaline-earth element compositions. After growth surface quality characterizations with *in situ* RHEED and *ex situ* atomic force microscope are shown in Figures S7 and S8. The slender reciprocal spots confirm the crystallinity of these cuprate thin films prepared by GOALL-Epitaxy (Figure 3e).

Figure 4 summarizes the parameter space covered by various oxide thin film techniques and the parameters required for different material systems. GOALL-Epitaxy exhibits oxidation power significantly surpassing conventional PLD by three orders of magnitude and OMBE by four, enhancing thermodynamic stability considerably. Their upper pressure limits are determined by evaporation mean free paths for OMBE, deposition rate of ablated materials for PLD, and availability of RHEED for GOALL-Epitaxy. Regarding lower temperature limits, while PLD provides higher kinetic energy to deposited materials by laser ablation compared to OMBE at lower evaporation temperatures, GOALL-Epitaxy provides higher kinetics by laser ablation within single atomic layer ranges during growth, more flexible for lower temperatures compared to single unit cell ranges for PLD. At higher temperatures, both GOALL-Epitaxy and OMBE are mainly limited by substrate heater capacity (laser heater versus typical resistive radiation), while PLD also considers heat dissipation at higher pressures. The distinct growth parameter regimes for nickelates (represented by $ReNiO_3$, where Re = rare earth), infinite-layer cuprates, and finite-layer cuprates (e.g. $La_2CuO_4$ and $YBa_2Cu_3O_{7-\delta}$) involves both thermodynamic and kinetic considerations. The near-vertical boundary at lower temperatures represents a kinetic limitation: for instance, more complex and larger unit cells in finite-layer cuprates (e.g. $YBa_2Cu_3O_{7-\delta}$) requires higher kinetic thresholds for lattice formation, comparing to infinite-layer cuprates. The boundary at higher temperatures is determined by thermodynamic constraints, with greater oxidation power providing increased thermodynamic stability at elevated growth temperatures [43].

Finite-layer cuprates can be grown more readily using conventional PLD or OMBE, thanks to their broad overlapping parameter spaces. However, the overlapping parameter spaces for nickelates and infinite-layer cuprates are significantly narrower, considerably limiting opportunities for optimization and the exploration of new phases and structures. In contrast, GOALL-Epitaxy's expanded parameter space offers significant advantages for the design and discovery of new materials. The superior oxidation capabilities support higher growth temperatures, which in turn enhance crystallinity by higher growth kinetics within single atomic layers. Additionally, GOALL-Epitaxy achieves atomic-layer precision—on par with OMBE and exceeding the unit cell precision in conventional PLD—thereby optimizing structural precision while preserving material versatility.

In conclusion, GOALL-Epitaxy marks a leap for thin film techniques. It not only amalgamates the strengths of both OMBE and PLD while overcoming their limitations, but also surpasses them substantially, particularly in oxidation power. GOALL-Epitaxy vastly expands the design scope of strongly correlated electron systems with tailored functionalities to previously uncharted parameter spaces, such as higher-$T_C$ superconductors, highlighting a transformative impact on the epitaxy of complex oxide materials.

## Methods

**Nickelate Ruddlesden-Popper (RP) phases growth.** Growth temperatures are typically set between 550 °C and 800 °C, with ozone chamber pressures ranging from $1 \times 10^{-5}$ mbar to 0.1 mbar. LaO$_x$ and NiO$_x$ targets are alternatively ablated using a KrF excimer laser ($\lambda$ = 248 nm, pulse duration 25 ns) for the sequential growth of different atomic layers. Stoichiometry control for various RP phases, encompassing the required pulse numbers for each atomic layer's completion and the laser energy, is initially calibrated using LaNiO$_3$ film synthesis and subsequently fine-tuned.

During deposition, typical the laser fluence on the $LaO_x$ and $NiO_x$ targets was ~1.4 $J/cm^2$ at 2 Hz and ~1.4 – 1.8 $J/cm^2$ at 2 Hz, respectively. The typical number of laser pulses was about 90 for each LaO layer and 100 for each $NiO_2$ layer. In case of Sr doping, $(La,Sr)O_x$ target was used instead of $LaO_x$, in which Sr ratio is around 0.2. All growth of thin films is monitored in real-time using 30 keV RHEED. Substrates mounted on a flag-type sample holder are heated by laser with highest temperature exceeding 1100 °C.

**Atomic hydrogen reduction.** The grown $(La,Sr)NiO_3$ films without any capping were *in situ* transferred under ultrahigh vacuum from the oxidation growth chamber into a dedicated reduction chamber. The atomic hydrogen were generated by a commercial hydrogen sources from Dr. Eberl MBE-Komponenten GmbH. For different levels of oxygen content, the reaction temperature ranges from 250 °C to 300 °C and the flux ranges from 0.5 to $3 \times 10^{15}$ atoms/$cm^2$s.

**Infinite-layer cuprate growth.** The infinite-layer cuprate films were grown on an (001)-oriented $SrTiO_3$, or 0.05% Nb doped $SrTiO_3$, or $NdGaO_3$ single-crystal substrate with a KrF excimer laser ($\lambda$ = 248 nm, pulse duration 25 ns). During deposition, the laser fluence on the ceramic $SrO_x$ targets was 1.2 $J/cm^2$ at 3 Hz, and on the ceramic $CuO_x$ and $CaO_x$ targets was 1.5 $J/cm^2$ at 3 Hz. The oxygen partial pressure was set to $1-2\times10^{-2}$ mbar, and the substrate temperature was maintained from 550 °C to 750 °C. The typical number of laser pulses was about 160 for each $CaO_x$ layer, 50 for each $SrO_x$ layer, and 60 for each $CuO_x$ layer in the growth of stoichiometric $CaCuO_2$, $SrCuO_2$, and $Sr_{0.5}Ca_{0.5}CuO_2$. After deposition, the $SrCuO_2$ and $Sr_{0.5}Ca_{0.5}CuO_2$ films were annealed at about $1\times10^{-7}$ mbar under 520 °C for 10 min, and the $CaCuO_2$ films were cooled down to room temperature at a rate of 10 °C/min at growth oxygen partial pressure. All growth of thin films is monitored in real-time using 30 keV RHEED. Substrates mounted on a flag-type sample holder are heated by laser with highest temperature exceeding 1100 °C.

**Target preparation.** $CaO_x$, $SrO_x$, and $LaO_x$ targets are reactive in ambient atmospheres, forming hydroxides upon contact with water, while $LaO_x$ additionally absorbs $CO_2$ from the air. To mitigate these reactions, these targets were sintered in a furnace within a glovebox under a dry Ar atmosphere. They were then mounted onto the target holders and rapidly transferred to the vacuum chamber through a load-lock system to minimize their exposure to air.

**Substrate preparation.** To achieve sharp step and terrace surfaces on $TiO_2$-terminated $SrTiO_3$ (001) substrates (Shinkosha, Japan), annealing processes were executed at 1100 °C for a protracted period of 6 hours within an air atmosphere. For the $LaAlO_3$ (001) substrates (MTI, China), an initial pre-treatment entailed immersion in boiling deionized water for 15 minutes. Subsequent annealing was performed under identical temperature, time and atmospheric conditions as $SrTiO_3$. The substrates were subjected to a repeated deionized water treatment-annealing protocol when needed, effectively yielding $AlO_2$-terminated $LaAlO_3$ substrates.

**X-ray diffraction (XRD).** Crystallographic characterization of thin-film specimens was performed by SmartLab, an automated multipurpose X-ray diffractometer, from Rigaku Corporation, encompassing theta-2theta scans and RSM.

**Scanning transmission electron microcopy (STEM).** STEM HAADF imaging of $La_3Ni_2O_7$ and $CaCuO_2$ was photographed using a FEI Titan Themis G2 at 300 kV, with a double spherical-aberration corrector (DCOR) and a high-brightness field-emission gun (X-FEG) with a monochromator is installed onto this microscope. The inner and outer collection angles for the STEM images ($\beta 1$ and $\beta 2$) were 48 and 200 mrad, respectively, with a semi-convergence angle of 25 mrad. The beam current was about 80 pA for high-angle annular dark-field imaging and the EDS chemical analyses. All imaging was done at room temperature. The cross-section STEM specimens of $La_3Ni_2O_7$ and $CaCuO_2$ were prepared using a FEI Helios 600i dual-beam FIB/SEM machine. Before extraction and thinning, we used electron beam-deposited platinum and ion beam-deposited carbon to protect the sample surface from ion beam damage. The cross-section STEM specimen of the 1212 structure was prepared using a Thermo Scientific Helios G4 HX machine, and was protected by electron beam-deposited platinum and ion beam-deposited carbon before extraction and thinning. The STEM annular bright field (ABF) and HAADF imaging of 1212 structure was photographed using a Thermo Scientific Themis Z at 200 kV with spherical-aberration corrector. And EDS data of 1212 structure were obtained using the Super X FEI System in STEM mode.

**Low-temperature transport measurements.** Electric transport measurements were performed in a closed-cycle helium-free system (base temperature below 1.5K). The four terminal electrical measurements were carried out through either the standard lock-in technique with an AC current of 1 μA (13.333 Hz) or a Keithley 6221 current source and 2182A voltmeter in a delta mode configuration.

**COMSOL simulation of ozone gas flow in the chamber.** A multiphysics numerical simulation, integrating fluid flow and heat conduction, was conducted using a coupled approach. The finite element method (FEM) was employed to solve the steady-state equations of fluid dynamics and heat transfer. The simulation's geometric model was designed to closely mimic the experimental growth apparatus, with dimensions of 400 mm for both the chamber's diameter and height, and a 70 mm gap between the PLD target and the heating stage. The computational domain was discretized using a free tetrahedral mesh. The simulation accounted for oxygen's properties, specifying the density, dynamic viscosity, and thermal conductivity accordingly. The temperature of the heating stage was maintained at 600 °C, while the chamber walls were kept at 20 °C. For the boundary conditions, the inlet was given a normal inflow velocity of 3 mm/s, while the outlet was treated as a pressure outlet set to 1 Pa.

## Supplementary Data

Supplementary data are available at NSR online.

## Acknowledgements

TEM characterization was performed at the Cryo-EM Center and Pico Center from SUSTech Core Research Facilities that receives support from the Presidential Fund and Development and Reform Commission of Shenzhen Municipality.

## Funding

This work was supported by the National Key R&D Program of China (No. 2022YFA1403100), the Natural Science Foundation of China (Nos. 92265112 and 12374455), and Guangdong Provincial Quantum Science Strategic Initiative (No. GDZX2401004, GDZX2201001 & SZZX2401001). J.L. and Q.Y. acknowledge the support from Guangdong Innovative and Entrepreneurial Research Team Program (Grant No. 2019ZT08C044) and Shenzhen Science and Technology Program (No. 20200925161102001). D.L. acknowledges the support from Guangdong Basic and Applied Basic Research Grant (No. 2023A1515011352) and Hong Kong Research Grants Council (CityU 21301221, CityU 11309622). Y.S. acknowledges the support from Natural Science Foundation of China (No. 12141402). G.M.Z. acknowledges the support of National Key Research and Development Program of China (Grant No. 2023YFA1406400).

## Author contributions

Q.K.X. and Z.C. supervised the project. Z.C. initiated the study and coordinated the research efforts. Z.C., H.H., and G.Z. designed instruments and experiments. G.Z., W.L., and Z.N. performed growth of nickelate thin films. H.H., F.W., and Y.L. performed growth of cuprate thin films. H.W. and C.D. performed reduction of nickelate films. H.W. and Z.N. performed low-temperature transport measurements. Q.Y. and J.L. provided STEM imaging. J.L. and C.Y. calculated the cuprate structures. Y.S., D.L., and all other authors participated discussions. G.M.Z. designed the 1212 nickelate material structure. Z.C., H.H. and G.Z. wrote the manuscript with input from all other authors.

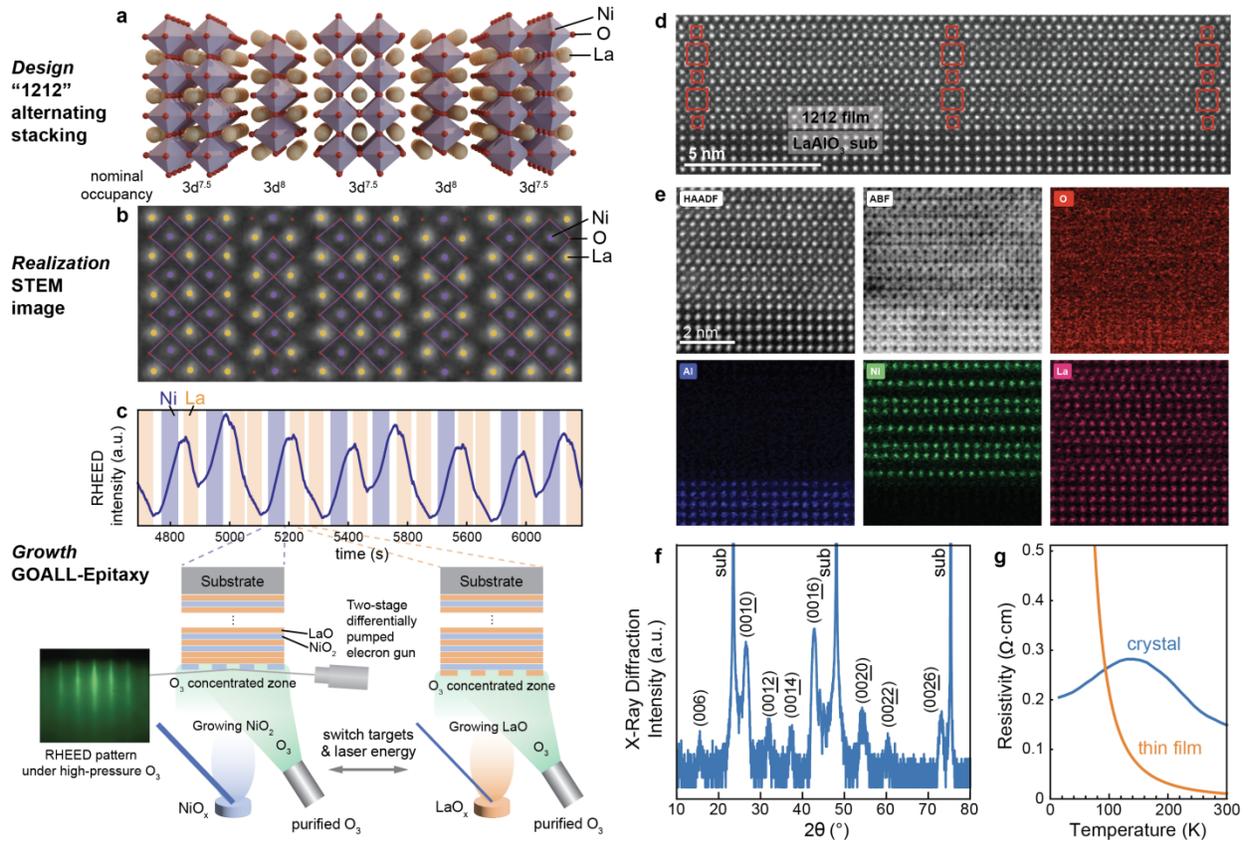

**Fig. 1 | Growth of a designed complex structure with gigantic-oxidative atomic-layer-by-layer epitaxy (GOALL-Epitaxy). a**, Schematic of a designed complex structure, featuring alternating stacking of single and double layers of $NiO_2$ (denoted as "1212"). This structure does not belong to the series of Ruddlesden-Popper phase. **b**, Magnified scanning transmission electron microscopy (STEM) image of the grown film, showing a region of lattice structure that is the same as that depicted in **a**. Atom positions are determined based on both high-angle annular dark field (HAADF) and annular bright field (ABF) images. **c**, Reflective high-energy electron diffraction (RHEED) intensity oscillations as a function of time. Blue and yellow blocks represent durations of $NiO_x$ and $LaO_x$ targets being ablated, respectively. Lower schematics describe how a complex structure is constructed atomic-layer-by-layer in the GOALL-Epitaxy setup. **d**, Larger-field-of-view HAADF image of the atomically sharp 1212 film and $LaAlO_3$ substrate interface. Red squares are guides to the eye for single-$NiO_2$ structures and double-$NiO_2$ structures. **e**, HAADF, ABF, and atomically resolved Energy-dispersive X-ray spectroscopy (EDS, for O, Al, Ni, and La, respectively) images of the same region of lattice. Alternating single and double layers of $NiO_2$ are exhibited. **f**, X-ray diffraction (XRD) in log scale of a 10-nm 1212 film. **g**, Resistivity as a function of temperature for a 20-nm 1212 film and a crystal.

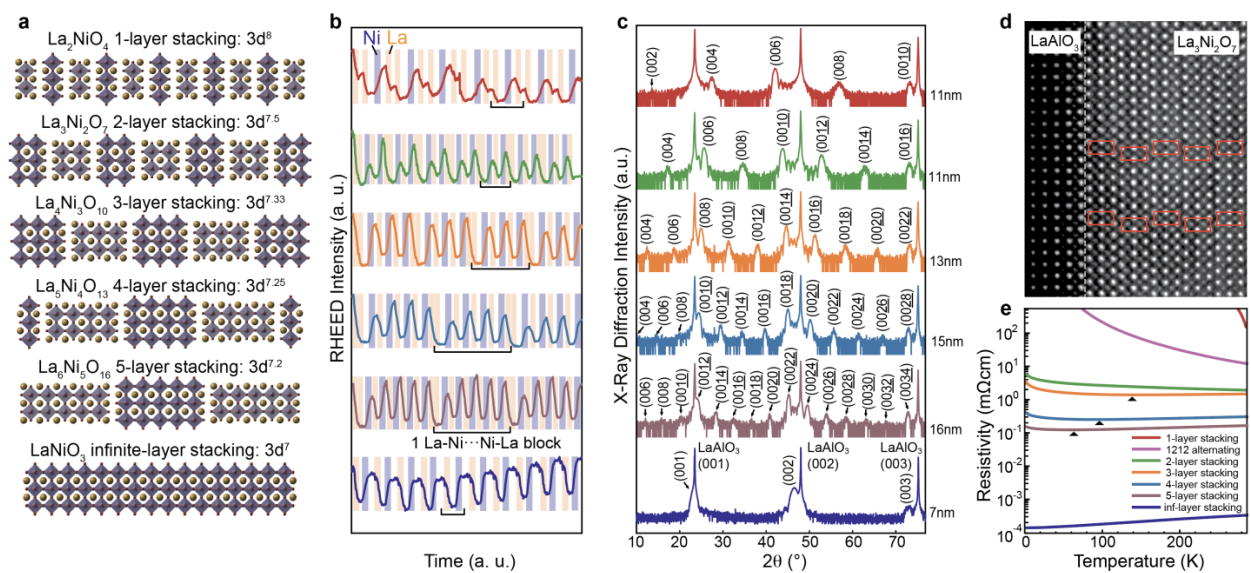

**Fig. 2 | Growth of complex nickelate structures. a**, Schematic structures of a series of Ruddlesden-Popper (RP) phases of nickelates. **b**, RHEED oscillations corresponding to each of the RP structures in a. Blue and yellow blocks represent durations of $NiO_x$ and $LaO_x$ targets being ablated, respectively. **c**, XRD of thin films corresponding to different RP structures. **d**, STEM HAADF image of a grown double-layer stacking structure, $La_3Ni_2O_7$. Red rectangles are guides to the eye highlighting $LaO$-$NiO_2$-$LaO$-$NiO_2$-$LaO$ blocks, as a fundamental unit that construct the structure. Dashed gray line indicate the interface. **e**, Resistivity-versus-temperature curves for various as-grown films without post-annealing. Possible oxygen loss may be present. Black filled triangles mark the points where resistivity begins to increase as the temperature decreases.

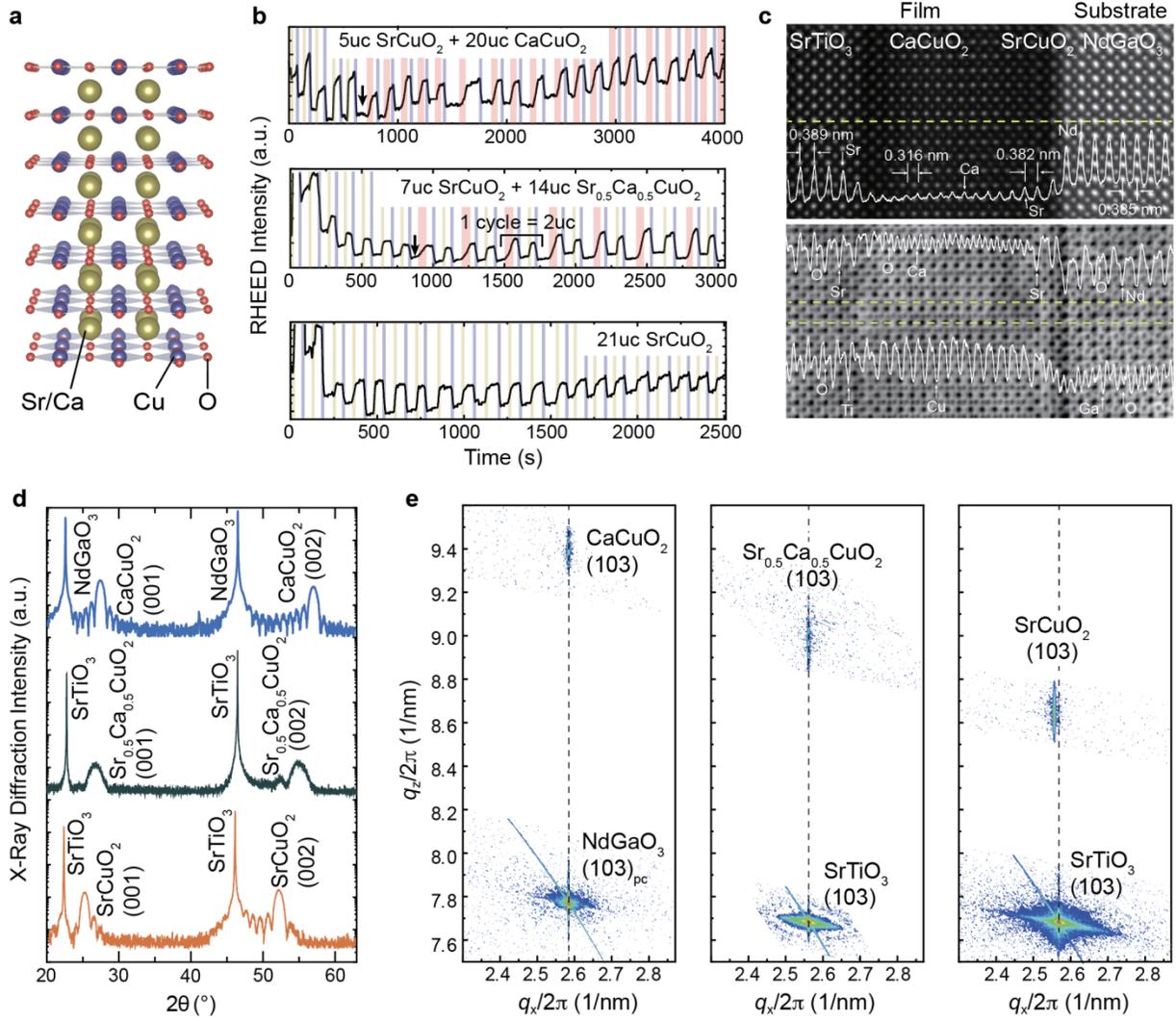

**Fig. 3 | Growth of infinite-layer cuprates. a**, Schematic lattice structure of infinite-layer cuprates. **b,** Example RHEED oscillations and patterns of $CaCuO_2$ growth with $SrCuO_2$ buffer on $NdGaO_3$ substrate (top), $Sr_{0.5}Ca_{0.5}CuO_2$ with $SrCuO_2$ buffer and $SrCuO_2$ growth on $SrTiO_3$ substrates (middle and bottom). Blue, pink, and yellow blocks represent durations of $CuO_x$, $SrO_x$, and $CaO_x$ targets being ablated, respectively. Note that 1 cycle of sequential depositions of $CaO_x$, $CuO_x$, $SrO_x$, and $CuO_x$, corresponding to 2UC $Sr_{0.5}Ca_{0.5}CuO_2$. **c**, STEM HAADF (top) and ABF (bottom) images of a $SrTiO_3/CaCuO_2/SrCuO_2/NdGaO_3$ sample, with inset showing the image intensity as a function of distance. **d** and **e**, XRD spectra along out-of-plane axis and reciprocal space mappings (RSM) of the three representative samples.

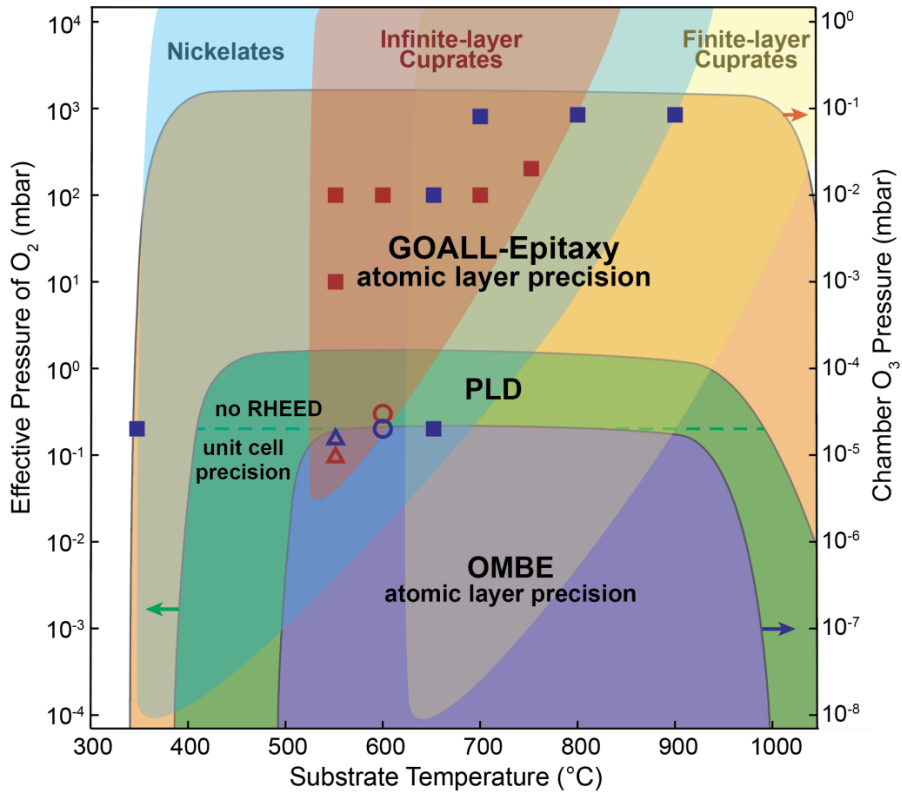

**Fig. 4 | New frontier in growth parameter space.** The parameter spaces covered by GOALL-Epitaxy, PLD, and OMBE techniques are represented by orange, green, and blue areas, delineated with solid grey lines, respectively. The parameter spaces where nickelates (here represented by ReNiO$_3$, where Re = rare earth), infinite-layer cuprates, and finite-layer cuprates (such as La$_2$CuO$_4$, YBa$_2$Cu$_3$O$_{7-\delta}$, etc.) can be grown are indicated by partially transparent blue, red, and yellow areas without border lines. For the PLD technique, the dashed green line distinguishes the regions where RHEED is applicable or not. Solid squares, empty circles, and empty triangles represents the example growth parameters by GOALL-Epitaxy, PLD [25,26], and OMBE [33,41-43], respectively. Blue and red symbols correspond to nickelates and infinite-layer cuprate growths, respectively.

# Supplementary Information

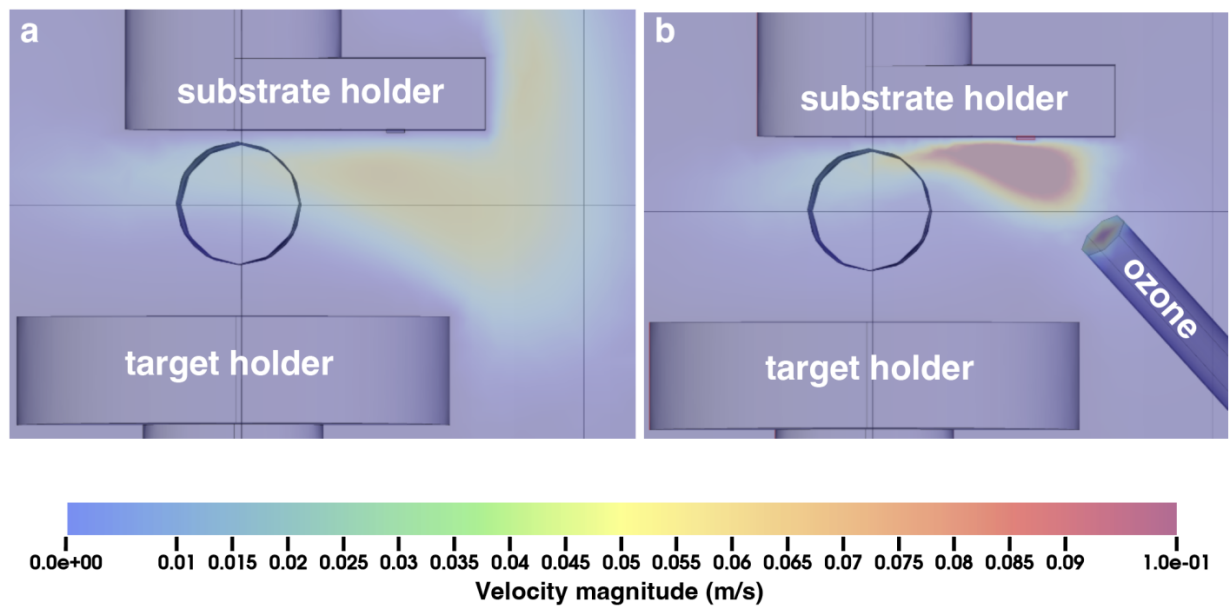

**Fig. S1 | COMSOL simulation of ozone gas flow in the chamber. a** and **b** depict the velocity magnitudes of ozone gas flow, with the nozzle opening situated 29 cm and 3.8 cm from the substrate, respectively. The velocity magnitudes at the substrate surface are 0.08 m/s for case a and 0.02 m/s for case b, indicating that positioning the nozzle closer to the substrate enhances the oxidation power by nearly half an order of magnitude. It is important to note that the oxidation power is predominantly determined by the initial contact of ozone molecules with the substrate due to their propensity for decomposition; therefore, we utilize velocity magnitude as a proxy for oxidation power rather than chamber pressure.

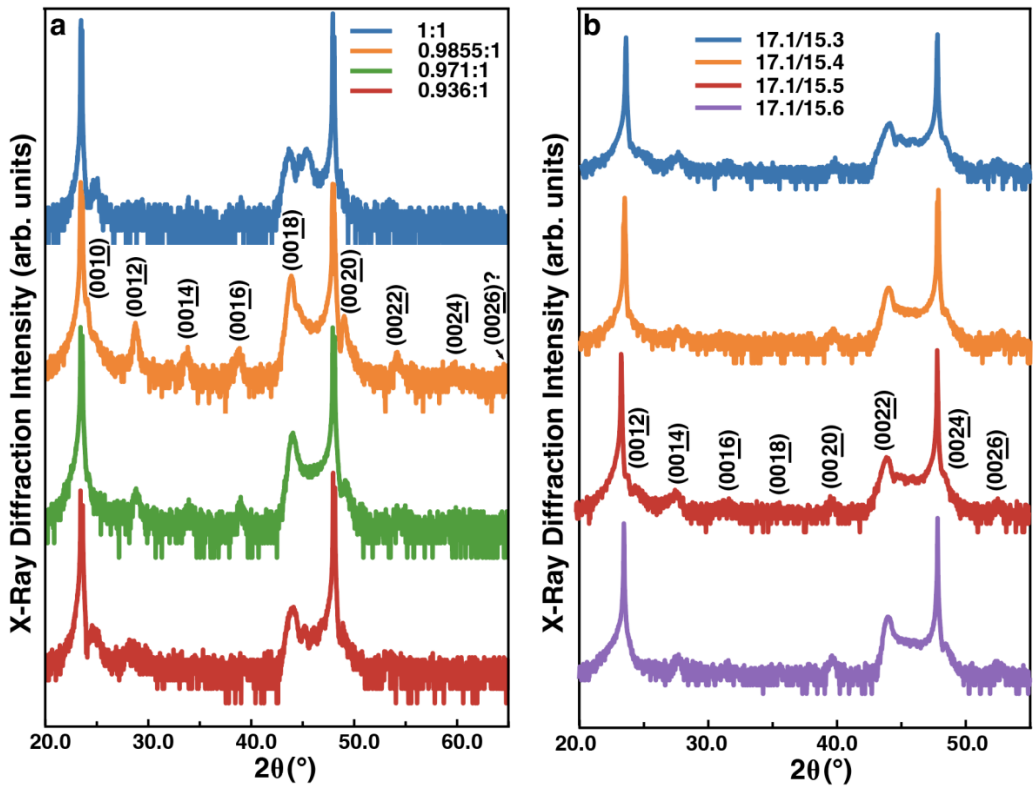

**Fig. S2 | Stoichiometry control.** In GOALL-Epitaxy, the stoichiometry can be tuned by varying the pulse number for each atomic layer (demonstrated in **a** for $La_5Ni_4O_{13}$), with the precision limit defined by single pulse of ablation, and by finer adjustment of the laser power (demonstrated in **b** for $La_6Ni_5O_{16}$). Different spectra in **a** represent different pulse number ratio between $LaO_x$ target and $NiO_x$ target ablations. Different spectra in and **b** correspond to different average laser energy of one pulse measured before entering the chamber in unit of mJ for $LaO_x/NiO_x$ targets. Thickness of the films are around 10 nm. The theoretical stoichiometry precision limit, estimated based on the lowest adjustable digital increment of laser energy, is about 0.1%. In experiment, the smallest stoichiometry change is at 1%-2% level in **a**, and about 0.3% in **b**.

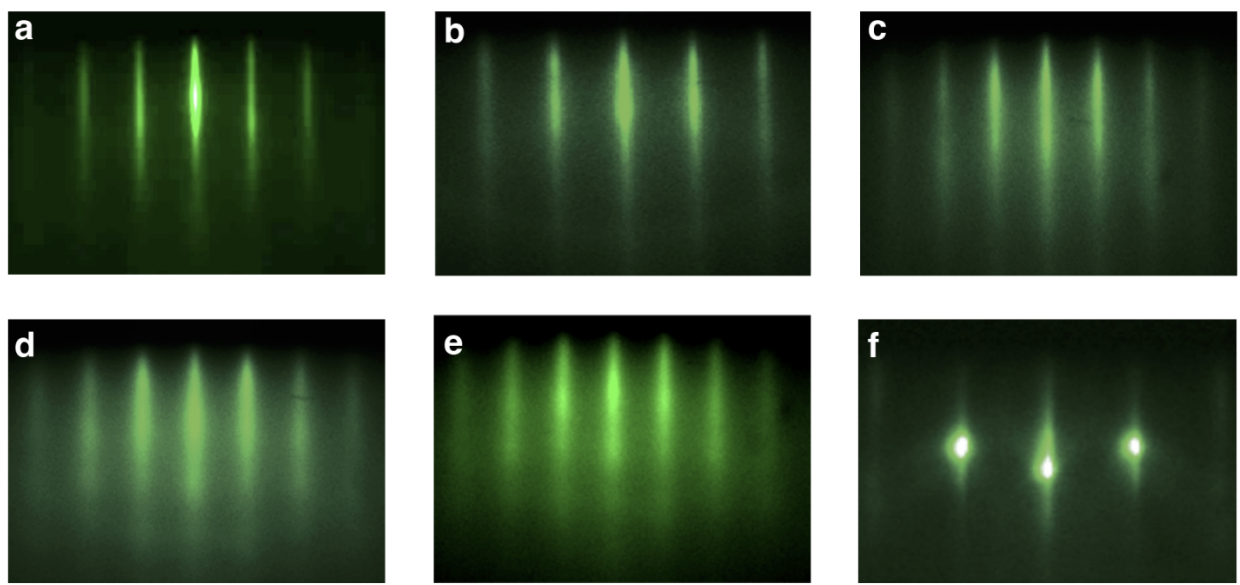

**Fig. S3 | Reflective high-energy electron diffraction patterns of nickelate Ruddlesden-Popper phases after growth. a-e**, one-layer to five-layer stacking structures $La_{n+1}Ni_nO_{3n+1}$, where $n$ is the number of consecutive layers in one stacking block. **f**, infinite-layer stacking structure $LaNiO_3$.

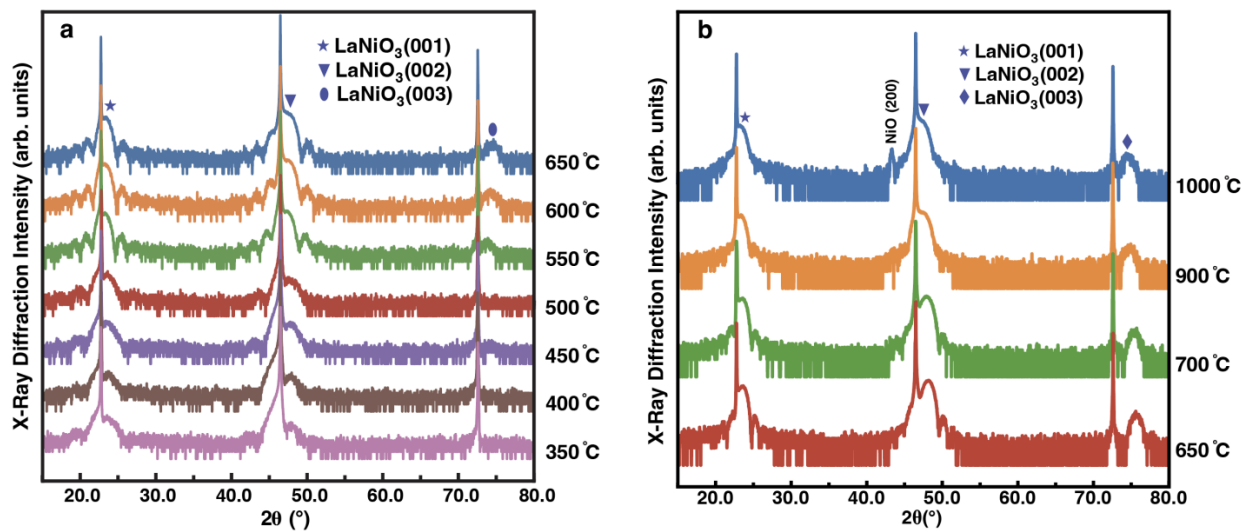

**Fig. S4 | Expansion of growth temperature range for LaNiO$_3$ on SrTiO$_3$ substrate.** GOALL-Epitaxy harnesses the kinetic energy of plasma from laser ablation for the top atomic layer being deposited, which enables the growth of single-crystalline epitaxial LaNiO$_3$ thin films at temperatures as low as 350 °C at $2 \times 10^{-5}$ mbar O$_3$ chamber pressure (**a**) and as high as 900 °C at 0.08 mbar O$_3$ chamber pressure (**b**). At 900 °C, LaNiO$_3$ remains marginally stable, but at 1000 °C, the emergence of a NiO peak suggests thermodynamic instability. GOALL-Epitaxy extends the lower growth temperature limit, signifying higher growth kinetics, and enhances the upper growth temperature, indicating improved thermodynamic stability, compared to prior methods.

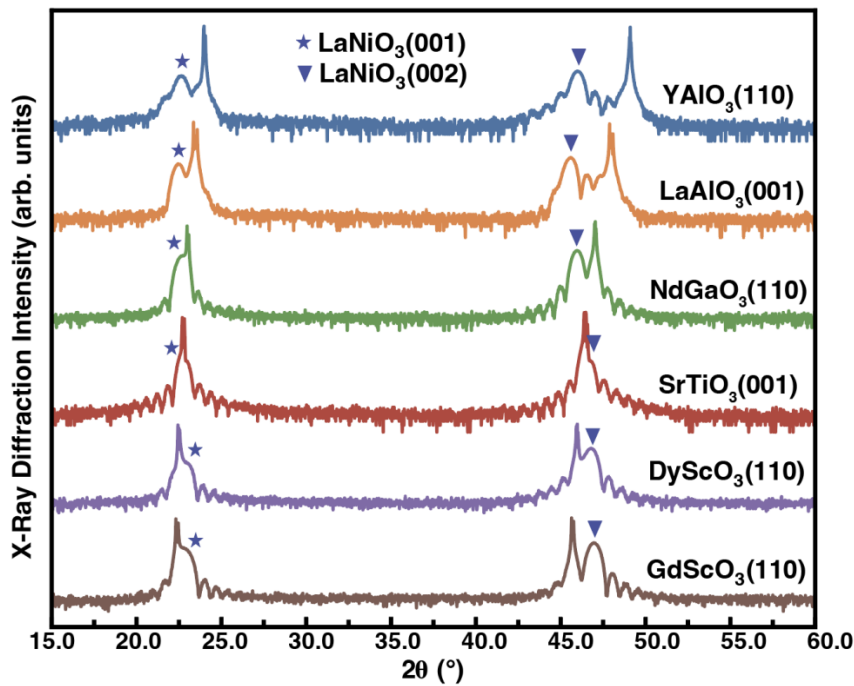

**Fig. S5 | Wide-range epitaxial strain tuning of LaNiO₃.** The quality of the films produced by GOALL-Epitaxy allows for a broad range of epitaxial strain across various substrates. The consistent shift in the out-of-plane lattice constant across films up to about 14 nm thick (grown at 650 °C and $2 \times 10^{-5}$ mbar O₃ chamber pressure), from GdScO₃ to LaAlO₃ substrates, demonstrates the coherent application of strain, amounting to approximately a 3% change in the in-plane lattice constant. Notably, on YAlO₃ substrates, the out-of-plane lattice constant is smaller than that on LaAlO₃, despite YAlO₃ having a smaller in-plane lattice constant, suggesting a relaxation of strain under the extreme high compressive conditions of YAlO₃.

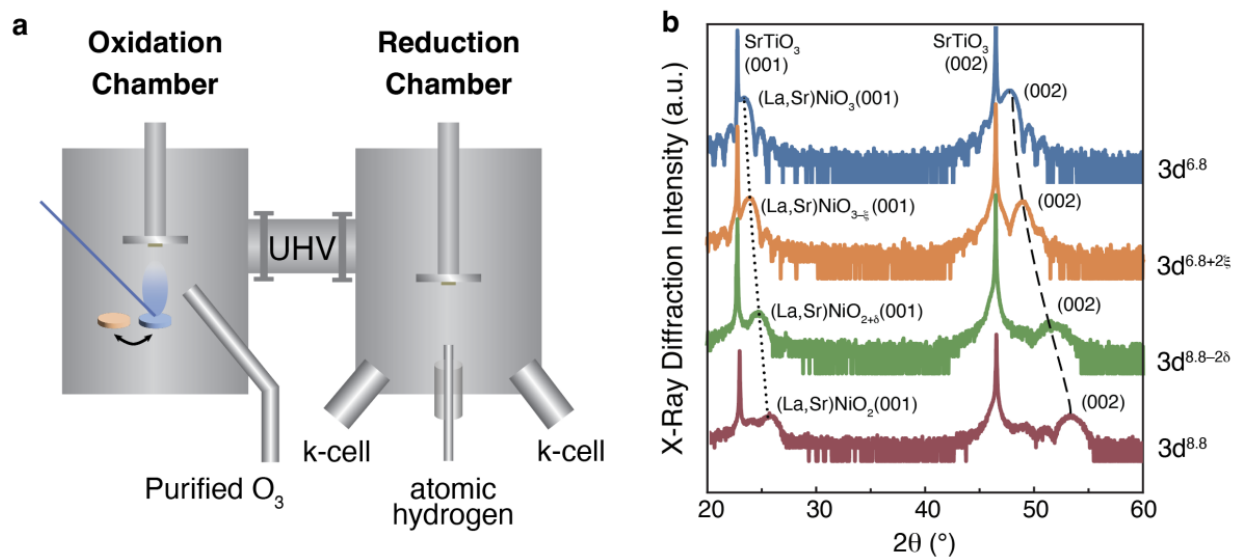

**Fig. S6 | Continuous valence tuning via *in situ* atomic hydrogen reduction. a**, Schematic diagram of the oxidation and reduction setup. K-cell: Knudsen effusion cell. UHV: ultrahigh vacuum. **b**, XRD data of a series of nickelates grown on SrTiO$_3$ substrate, having coherent infinite-layer structure but with different 3*d*-obital occupancy, tuned by varied oxygen content.

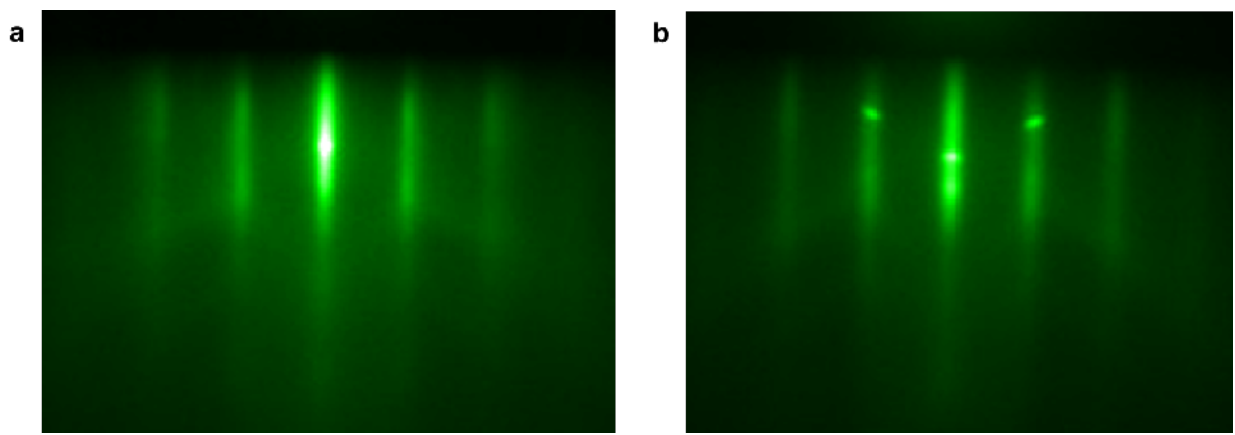

**Fig. S7 | RHEED patterns of cuprates after growth. a**, 20 unit-cell $SrCuO_2$ film on $SrTiO_3$ substrate. **b**, 20 unit-cell $CaCuO_2$ on $NdGaO_3$ substrate with 4 unit-cell $SrCuO_2$ buffer.

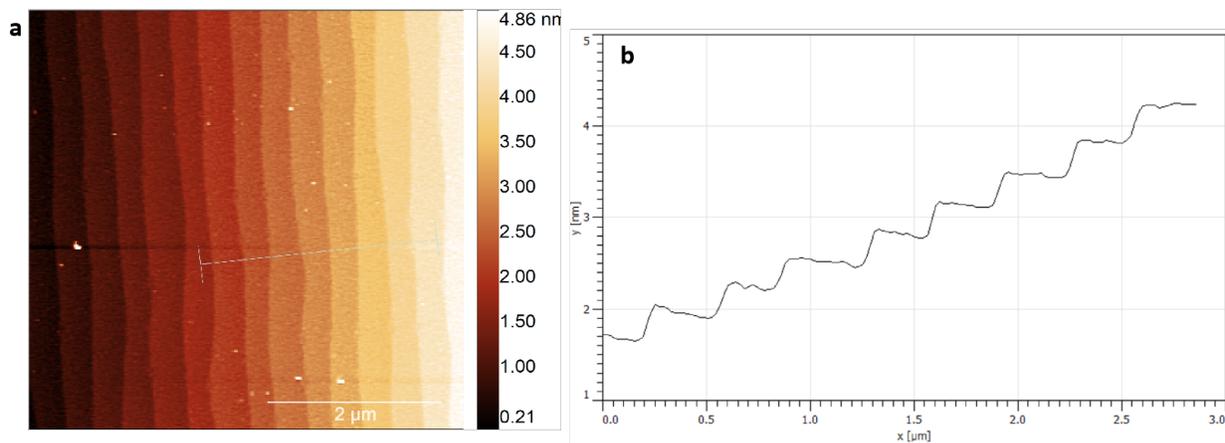

**Fig. S8 | Atomic force microscope topographic image of the CaCuO$_2$ film. a**, Atomically flat surface of 15 unit-cell CaCuO$_2$ film on NdGaO$_3$ substrate with 4 unit-cell SrCuO$_2$ buffer. **b**, The atomic steps of CaCuO$_2$ film surface.